# F5C-finder: An Explainable and Ensemble Biological Language Model for Predicting 5-Formylcytidine Modifications on mRNA

Guohao Wang[a,b], Ting Liu[c], Hongqiang Lyu[d] and Ze Liu[a,b*]

[a]College of Water Resources and Architectural Engineering, Northwest A&F University, Yangling, 712100, Shaanxi, China.

[b]Key Laboratory of Agricultural Soil and Water Engineering in Arid and Semiarid Areas, Ministry of Education, Northwest A & F University, Yangling, 712100, Shaanxi, China.

[c]Department of Mechanical Engineering, Faculty of Engineering, The University of Hong Kong, 999077, Hong Kong, China.

[d]School of Automation Science and Engineering, Faculty of Electronic and Information Engineering, Xi'an Jiaotong University, Shaanxi 710049, China

**\*Corresponding Author**

**Email:** Guohao Wang: guohaowang@ieee.org Ting Liu: tsing_liu_nwafu@163.com Hongqiang Lyu: hongqianglv@mail.xjtu.edu.cn \*Ze Liu: zeliu@nwafu.edu.cn

**Abstract:** As a prevalent and dynamically regulated epigenetic modification, 5-formylcytidine (f5C) is crucial in various biological processes. However, traditional experimental methods for f5C detection are often laborious and time-consuming, limiting their ability to map f5C sites across the transcriptome comprehensively. While computational approaches offer a cost-effective and high-throughput alternative, no recognition model for f5C has been developed to date. Drawing inspiration from language models in natural language processing, this study presents f5C-finder, an ensemble neural network-based model utilizing multi-head attention for the identification of f5C. Five distinct feature extraction methods were employed to construct five individual artificial neural networks, and these networks were subsequently integrated through ensemble learning to create f5C-finder. 10-fold cross-validation and independent tests demonstrate that f5C-finder achieves state-of-the-art (SOTA) performance with AUC of 0.807 and 0.827, respectively. The result highlights the effectiveness of biological language model in capturing both the order (sequential) and functional meaning (semantics) within genomes. Furthermore, the built-in interpretability allows us to understand what the model is learning, creating a bridge between identifying key sequential elements and a deeper exploration of their biological functions.

## 1. Introduction

With the recent advances in epigenetics, more than 170 distinct modifications have been identified on RNA. Among them, 5-formylcytidine (f5C) has emerged as an abundant regulated modification. f5C was first discovered in mitochondrial tRNA$^{Met}$ of bovine and nematode in 1994 [1,2] and was also found in squids, frogs, chickens, rats, and fruit flies in recent decades [3–5]. During the setting process of f5C on the transcriptome, C34 (position 34 of the mammalian mitochondrial methionine transfer RNA) is methylated to form 5-methylcytosine (m5C) under the catalysis of NSUN (NOL1/NOP2/sun domain) RNA methyltransferase. Subsequently, m5C is oxidized to 5-hydroxymethylcytosine (hm5C) and then to f5C [6]. Although the functions of f5C are still largely unknown, it strongly indicates that loss of f5C will result in pathological consequences [7]. Takemoto et al. found that f5C in mitochondrial tRNA$^{Met}$ plays a crucial role in recognizing and decoding the nonuniversal AUA codon as Met [8]. In addition, two pathogenic point mutations in mitochondrial tRNA$^{Met}$ were found to prevent NSUN3 (NOP2/Sun RNA Methyltransferase 3)-mediated methylation. Without NSUN3, mitochondrial protein synthesis will dramatically decrease and reduce oxygen consumption, leading to defective mitochondrial activity [7]. Murakami et al. found that f5C is essential for mice's embryonic development and respiratory complexes [9]. However, many functions of f5C are still unknown, such as the contribution of f5C to the structure of the hmtRNA$^{Met}$ and its possible participation in either chain initiation or chain elongation by this unique tRNA$^{Met}$ [10].

To make accurate identification of f5C, researchers have proposed several approaches based on biological experiments up to now. On the basis of Friedlander synthesis, a bisulfite-free and single-base-resolution method was developed by inducing the transition of f5C to T transition [11,12]. However, this method is limited in application due to no way to efficiently and completely convert m5C to f5C [13]. To address this limitation, Liu et al. present an alternative method, named TET-assisted pyridine borane sequencing (TAPS), for the detection of m5C and hm5C. TAPS fusions TET oxidation to 5-carboxycytosine (5cac) with pyridine borane reduction of ca5C to Dihydrouracil (DHU), and DHU will be converted to thymine during Polymerase Chain Reaction (PCR) [14]. Compared with bisulfite sequencing, TAPS achieved better performance in whole-genome sequencing of mouse embryonic stem cells [15]. Inspired by their approach, Wang et al. developed a mutation-assisted profiling method in 2022 [16]. This mutation-assisted profiling method, named f5C-seq, can provide a single-base resolution map of f5C on the transcriptome.

Although experiments can provide reliable location information on mRNA, it takes considerable time and cost for the verification of all the f5C candidates. Recently, with the development of machine learning, computational methods have become increasingly popular as useful alternative methods [17-20]. To date, several computational methods have been developed for the recognition of RNA or DNA modification. For example, iDNA6Ma-Rice [17], iDNA6mA-PseKNC [18], i6mA-Fuse [19], Meta-i6mA [20], and i6mA-vote [21] are excellent prediction models for the N6-methyladenosine (6mA) modification. For the N7-Methylguanosine (m7G) and m5C prediction, m7Gpredictor [22], iRNA-m7G [22], BERT-m7G [23], Staem5 [24], and m5C-HPromoter [25] also are suitable models for their prediction task. Although the above models achieved excellent performances for the specific recognition tasks, the recognition model for the f5C modification has not been reported yet. Therefore, there is a strong motivation to establish a model for f5C identification.

Deep learning has demonstrated remarkable success in natural language processing (NLP), leading to the emergence of numerous language models [26–29], including long short-term memory (LSTM) networks [30,31] and transformers with multi-head attention mechanisms [32]. Large language models can now represent complex relationships between words, exhibiting improved generalization and robustness [33]. Inspired by the success of large language models, researchers have proposed biological language models, which have demonstrated promising results in bio-sequence prediction tasks [34–36]. Given the similarities between sentences and bio-sequences [37], biological language models hold the potential for predicting modifications in DNA or RNA. For example, PlncRNA-HDeep [38] employs a combination of LSTM and convolutional neural networks (CNN) for plant long noncoding RNA identification. EMDLP [39] utilizes one-hot encoding with ensemble dilated convolution and Bidirectional LSTM (BiLSTM) to identify post-transcriptional RNA modifications. BERT-m7G [23], a fusion model based on Bidirectional Encoder Representations from Transformers (BERT) [29], was designed for N7-Methylguanosine site recognition. DNA4mC-LIP predicts 4-methylcytosine (4mC) by integrating predictions from six existing models through ensemble learning, while DeepTorrent leverages inception modules and attention mechanisms to enhance 4mC prediction accuracy. For 5mC site detection, iPromoter-5mC combines predictions from multiple models using fully connected layers, and BiLSTM-5mC extracts features from nucleotide property and frequency encoded sequences via BiLSTM for 5mC prediction. Compared to statistical learning models, these models require fewer manual resources for feature selection and optimization while extracting more comprehensive and deeper hidden features. The growing number of research exploring the potential of biological language models in DNA or RNA methylation prediction has yielded significant progress in improving predictive performance. However, existing predictors have not fully harnessed the power of feature representation learning, particularly in discovering key sequential patterns crucial for elucidating RNA methylation mechanisms. This results in deep learning models with poor interpretability, hindering the identification of important sequence-based influences in RNA methylation prediction. Moreover, existing approaches leave several key questions unanswered: (1) What specific knowledge does the biological language model extract from the sequence? Does it contain additional discernible information that can guide RNA methylation development? (2) From a computational perspective, what is the most significant feature of f5C modification? (3) Furthermore, there is a lack of efficient identification models for f5C modification, as well as a lack of interpretable methods to characterize its features.

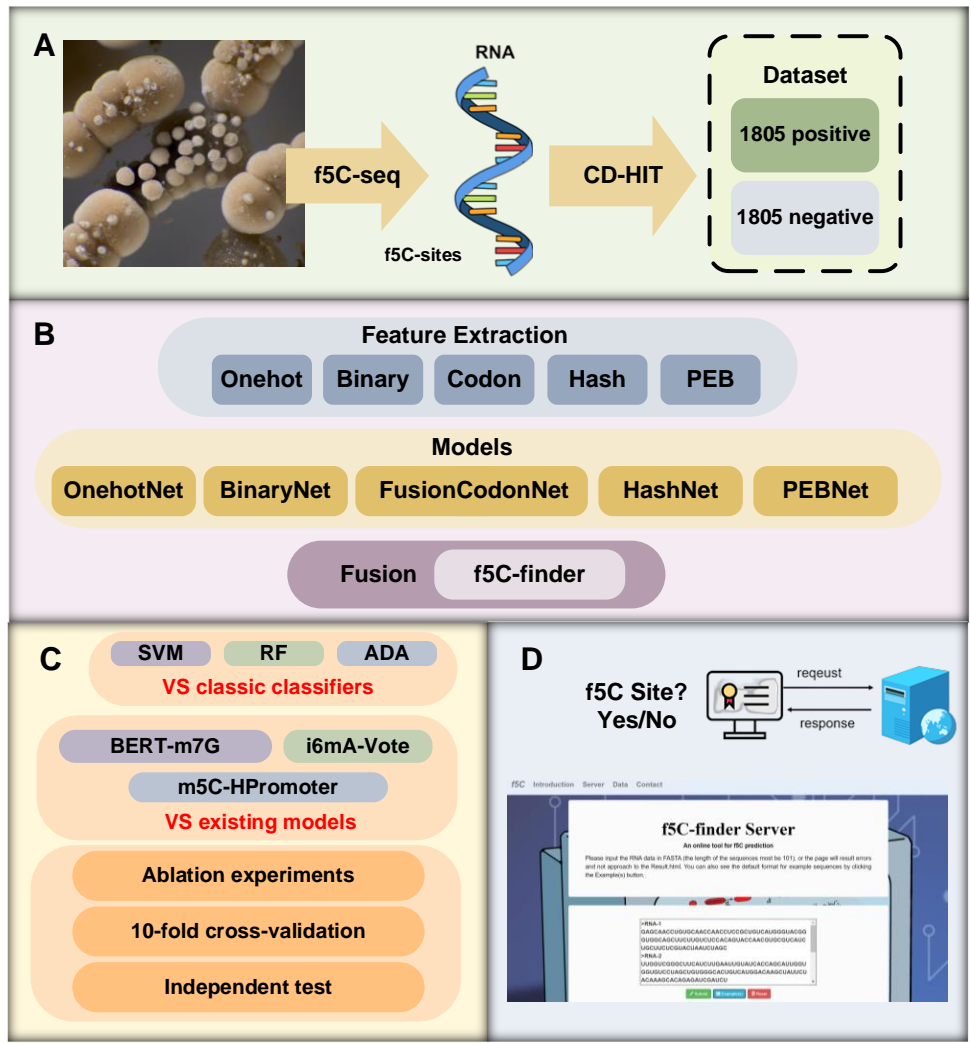

**Figure 1.** The workflow of f5C-finder. (A) Dataset formation. (B) Feature extraction and model construction. (C) Model evaluation and comparison. (D) Webserver implementation of f5C-finder.

Motivated by these challenges and the lack of existing models for identifying f5C modification, this study proposes f5C-finder, an explainable ensemble biological language model for f5C identification. The workflow is shown in Figure 1. Firstly, RNA sequence with the f5C modification was obtained and the dataset with an equal number of positive and negative samples was formatted. Secondly, five feature schemes were utilized for the feature extraction of RNA sequences, and combining the language model with LSTM and multi-head attention mechanism, 5 different neural network-based models were separately developed. The hyperparameters of these models were determined through ablation experiments. Thirdly, f5C-finder was established through the ensemble learning method. After that, to evaluate the performance of f5C-finder and demonstrate the superiority of this model, a comparative analysis with six ensemble learning models was conducted using 10-fold cross-validation and independent testing. The six ensemble models include three classic classifiers which were separately developed with the five features and integrated with the same ensemble learning method and three presented models. They were ensemble Random Forests (RF) [40], ensemble Support Vector Machine (SVM) [41], ensemble Adaboost (ADA) [42], BERT-m7G, i6mA-Vote, and m5C-HPromoter. Finally, to enhance user experience and facilitate the use of f5C-finder, a web server was also developed and can be accessed through http://f5c.m6aminer.cn/.

## 2. Material and methods

### 2.1. Original data acquisition

The single-base-resolution site information of f5C was collected from the reference [16], which identified 1,892 f5C sites on the whole transcriptome of Saccharomyces cerevisiae by using f5C-seq. In this study, these single-base-resolution sites were mapped to the genome of Saccharomyces cerevisiae, and the 50bp sequences centered with the f5C sites were intercepted as the positive samples. To generate the negative samples, the sequences of equal length with cytosine in the middle, which has no repetition with the positive samples, were intercepted as the negative samples. To remove the redundant sequences from the dataset, the CD-HIT-EST software [43] was adopted and the strictest threshold was set to 0.8. Finally, 1,805 positive and 1,805 negative non-redundant samples were obtained in total. Among them, 1,444 positive samples and 1,444 negative samples were included in the training dataset while 361 positive samples and 361 negative samples were included in the independent testing dataset.

### 2.2. Feature extraction

Due to the inability to directly put nucleotide sequences into the model, five feature extraction methods were employed in this paper. The 5 feature extraction methods are shown as follows:

*2.2.1. Binary code*

Binary code (Binary) is a simple but effective feature extraction algorithm to identify the origins of the replication sites [44] and 6mA modifications [17], which can be employed to convert each nucleotide into a binary vector. In this paper, a three-digit code was employed to represent a type of nucleic acid. An RNA sequence of 101 nucleotides can be represented mathematically as follows:

$$Binary = (b_1, b_2, b_3, b_4, \cdots, b_n, \cdots, b_{101}), b_n \in \begin{cases} A: [1,1,1] \\ C: [0,0,0] \\ G: [1,0,0] \\ U: [0,1,0] \end{cases} \quad (1)$$

Where $b_n$ is the nucleotide at position, $n$, and Binary maps each sequence into a matrix of 101×3.

*2.2.2. Hash decimal conversion*

Hash decimal conversion (Hash) is an effective feature extraction method by converting quaternary numbers into decimals [45]. For an RNA sequence, the two adjacent bases can be regarded as a two-digit-quaternary number. Firstly, a mapping of the four nucleotides is established as $h_A: 0, h_G: 1, h_C: 2, h_U: 3$. After that, the algorithm for converting quaternary to decimal can be expressed using the following formula:

$$s = \sum_{i=1}^{k} 4^{k-i} * h_i \quad (2)$$

Assuming that the length of a sequence is $m$ and the length of the selected segment is $k$, the number of decimal numbers obtained is $m - k + 1$. In this paper, the RNA sequence contained 101 nucleotides, and the $k$ was set to 2, so a 100-dimensional vector was generated using Hash for each sequence.

*2.2.3. Codon*

For different RNA sequences with f5C, the distribution of nucleotides is different. Thus, Codon was defined to extract features from this phenomenon. The encoding length of Codon is 5, among them, the first two values were generated from the single nucleic acid around central cytosine, based on the probability of nucleic occurrence in the f5C samples. The other 3 values represent the frequency of specific trinucleotide patterns containing f5C information. The central cytosine is in position 51, so based on the probability of trinucleotide from position 48 to 50, 50 to 52, and 51 to 53, the RNA sequences are separately expressed by 0.75, 0.5, 0.25, and 0 for the 5 values. The corresponding analysis is shown in Section. 4.1.

*2.2.4. One-hot encoding*

One-hot encoding (Onehot) [46] is a feature extraction method to convert nucleotides into a numerical representation. For the four nucleotides, each nucleotide is represented using a binary vector of length 4, where only one element is 'hot' (set it to 1) and the other elements are 'cold' (set it to 0). The position of the hot element corresponds to the position of the nucleotide in the encoding scheme. In this study, mapping is expressed as follows:

$$Onehot = (O_1, O_2, O_3, O_4, \cdots, O_n, \cdots, O_{101}), O_n \in \begin{cases} A: [1,0,0,0] \\ G: [0,1,0,0] \\ C: [0,0,1,0] \\ U: [0,0,0,1] \end{cases} \quad (3)$$

Where $O_n$ is the nucleotide at position, $n$. An RNA sequence of 101 nucleotides can be represented by a 101×4 matrix using One-hot encoding.

*2.2.5. Position encoding*

Position encoding for bio-sequence (PEB) [47,48] is a method to encode A, G, C, and U with a number from 1 to 4, the mapping is as follows:

$$PEB = (P_1, P_2, P_3, P_4, \cdots, P_n, \cdots, P_{101}), P_n \in \begin{cases} A: 1 \\ G: 2 \\ C: 3 \\ U: 4 \end{cases} \quad (4)$$

Where $P_n$ is the nucleotide at position, *n*. Given an RNA sequence of 101 nucleotides, PEB generates a corresponding 101-dimensional vector. PEB not only contains information on different kinds of nucleotides and their position in the sequences, but also keeps the relation between different nucleotides.

### 2.3. Biological language model construction

In this paper, five biological language models were constructed for the identification of f5C modification, among which OnehotNet and BinaryNet are based on LSTM, while PEBNet, HashNet, and FusionCodonNet are founded on the multi-head attention mechanism. By integrating these five biological language models, the f5C-finder was developed.

### 2.3.1. OnehotNet and BinaryNet

LSTM networks, renowned for capturing long-range dependencies within sequential data, offer significant advantages for modeling RNA sequences. Their unique cell structure allows them to retain information from earlier time steps, mitigating the vanishing gradient problem that plagues traditional language models. This capability is crucial for capturing the complex relationships and interactions between nucleotides within RNA sequences, which often influence their secondary structures and functions[38].

In this study, two biological language models leveraging LSTM layers were constructed, they are BinaryNet and OnehotNet. These models were designed to exploit the strengths of LSTM in modeling nucleotides, to accurately predict the f5C modification. Since the Binary and Onehot features preserve the positional information of the nucleic, using a language model architecture such as LSTM can effectively model the positional relationship of the nucleic, while also being conducive to obtaining a neural network model with better performance based on semantic information. The forward propagation process for both models is described below.

Firstly, Onehot feature and Binary feature generate the original encoding matrix, as shown in equation (5) and (6).

$$f_{enc1}(X) = Onehot(X) \tag{5}$$

$$f_{enc2}(X) = Binary(X) \tag{6}$$

While the feature inputs for the two models differ, their underlying structures are identical. For each sample, $X$, the 2 encoders are defined as $f_{enci}(X)$.

$$X' = Dropout(Flatten(tanh(LSTM(f_{enci}(X))))) \; i \in (1,2) \tag{7}$$

Where $X'$ represented the data in the middle process. The Flatten layer is added to convert the output of an LSTM layer into a one-dimensional vector that can be processed by Dropout layers. With the Dropout layer, the output of the Flatten layer is dropped out by the probability to prevent overfitting.

$$Output = Sigmoid(Dense(tanh(Dense(X')))) \tag{8}$$

As shown in equation (8), dense layers serve as the output layers. The sigmoid activation function constrains the model's output between 0 and 1 during forward propagation. Using a threshold of 0.5, outputs exceeding this value indicate sequences centered with the f5C modification, while the outputs below 0.5 indicate the absence of such modification.

### 2.3.2. Models with muti-head attention mechanism

Biological language models with multi-head attention mechanism are powerful tools for modeling nucleotide information and identifying RNA modifications. They can capture molecular information [49] with high accuracy, integrate different data sources [50] with robustness and generalizable ability, and provide interpretable results[51] for peptide secondary structure prediction. Thus, based on Codon, Hash, and PEB, this study develops three multi-head attention models for f5C modification prediction. These models leverage different aspects of sequence information to achieve accurate and interpretable predictions. There are two models with single-feature input, PEBNet and HashNet, and one model with multi-feature input, FusionCodonNet. The three features ($f_{enc1}(X)$ to $f_{enc3}(X)$) are defined as follows:

$$f_{enc3}(X) = PEB(X) \tag{9}$$

$$f_{enc4}(X) = Hash(X) \tag{10}$$

$$f_{enc5}(X) = FusionCodonNet(X) = \{Hash(X) + Codon(X) + PEB(X)\} \tag{11}$$

Following feature extraction via equations (9) to (11), the resulting feature vector is passed through an embedding layer, and subsequently, the attention layer.

$$X_{trans} = Attention\left(Embedding\left(f_{enci}(X)\right)\right) \ i \in (3,4,5) \tag{12}$$

$X_{trans}$, the output of the multi-head self-attention core, captures the positional information of each nucleotide through the attention mechanism. This allows the model to learn the semantic information of f5C during training. Subsequently, the output of the neural network is fed into equation (13) to obtain the final result.

$$Pre_{model} = Sigmoid(Dense(Dropout(GlobalMaxPooling(X_{trans})))) \tag{13}$$

Where GlobalMaxPooling layer, Dropout layer, Dense layer, and *Sigmoid* are utilized during forward propagation. After that, the prediction results for f5C, $Pre_{model}$, are generated. With the same threshold, 0.5, recognition results for f5C can be obtained.

### 2.3.3. Build a fusion model with an ensemble learning method

Ensemble learning is a machine learning method that combines predictions from multiple models to enhance overall prediction accuracy and robustness. By applying the ensemble learning method to the above five models, f5C-finder is developed. In this paper, the average of the five model outputs is the final predicted probability of the fusion model. Therefore, the threshold is also set to 0.5. To facilitate a comparative analysis between f5C-finder and traditional models, highlighting the advantages of language model architectures in the f5C modification recognition, RF, SVM, and ADA are separately developed with the five features and integrated with the same ensemble learning method. The hyperparameters for these models were initially searched using the Grid Search (Scikit-learn package) [52]. Then, the models with the optimal hyperparameters were utilized for ensemble learning. A linear fusion method was adopted for all the above models in this paper, as shown in equation (14).

$$Pre_{model} = (Pre_{OnehotNet} + Pre_{BinaryNet} + Pre_{PEBNet} + Pre_{HashNet} + Pre_{FusionCodonNet})/5 \qquad (14)$$

where $Pre_{model}$ represents the prediction results from *model* with ensemble learning method, and $Pre_{Onehot}$ represents the prediction results from the model with Onehot feature, and the other symbol representations are the same as above.

## 3. Performance evaluation

To evaluate the performance of f5C-finder and make comparison with different models, sensitivity (SN), specificity (SP), harmonic mean of sensitivity and specificity (F1-score), accuracy (ACC), and Matthews Correlation Coefficient (MCC) are utilized to evaluate the performance of different models, they are defined as follows:

$$SN = \frac{TP}{TP + FN} \qquad (15)$$

$$SP = \frac{TN}{TN + FP} \qquad (16)$$

$$ACC = \frac{TP + TN}{TP + TN + FP + FN} \qquad (17)$$

$$F1 = 2 \times \frac{\frac{TP}{TP+FP} \times \frac{TP}{TP+FN}}{\frac{TP}{TP+FP} + \frac{TP}{TP+FN}} \tag{18}$$

$$MCC = \frac{TP \times TN - FP \times FN}{\sqrt{(TP+FN)(TP+FP)(TN+FP)(TN+FN)}} \tag{19}$$

where *TP, TN*, *FN* and *FP* each represents true positive, true negative, false negative and false positive, respectively. Furthermore, the recognition performances are characterized by the Receiver-Operating characteristic (ROC) and Precision-Recall (P-R) curve. These metrics are complemented by the calculation of the Area Under the Curve (AUC) for the ROC and the corresponding Area Under the Precision-Recall Curve (AUPR), providing a comprehensive evaluation of the models' predictive accuracy. To evaluate the performance and make a comparison of different models, 10-fold cross-validation (10-CV) was conducted. In each fold, one subset served as the validation set while the remaining nine subsets were used for training. The average performance across all ten folds was calculated to obtain the final performance estimates on the six metrics. This approach provides a robust and unbiased estimate of the models' performance.

## 4. Results and discussion

### *4.1. Analysis of motif around f5C sites*

The sequences surrounding the f5C sites, as illustrated in Figure 2, display notable conservation. The motif "TTATTT" is most enriched at positions 45~50nt upstream of f5C, while the motif "TAAGA" shows the highest enrichment at positions 52~56nt downstream of f5C. Furthermore, the probability of cytosine occurrence at positions 50 and 52 is notably low. Consequently, Codon-based feature extraction was utilized in this study.

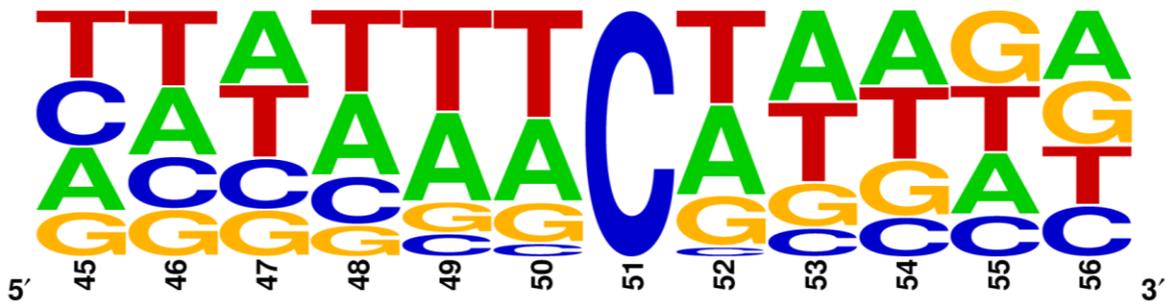

**Figure 2.** Sequence distribution around the f5C sites.

At the same time, Binary, Onehot, PEB, and Hash are fundamentally position mappings, which means the mapping of nucleotides to numbers is based on the position of elements in the sequence. Therefore, the above four feature extraction methods can locate cytosine at position 51, and establish effective statistics containing positional information for nucleotides in its immediate vicinity. This facilitates the biological language model in enhancing the final recognition performance through positional information. The learning effects of the biological language model on these features are demonstrated in Section 4.4.

### *4.2. Determine the hyperparameters of the neural networks*

Hyperparameters significantly influence the final recognition performance of neural network-based models. In this study, a 10-fold cross-validation strategy was employed on the training dataset for hyperparameter tuning of the five biological language models: OnehotNet, BinaryNet, HashNet, PEBNet, and FusionCodonNet. The architectures of BinaryNet and OnehotNet, both utilizing LSTM layers, are depicted in Figure 3(A). Two tuned hyperparameters were dropout rate of the Dropout layer and number of Dense layers. The AUC of OnehotNet and BinaryNet under various hyperparameter settings are illustrated in Figures 4(A) and 4(B), respectively. OnehotNet achieved the best AUC of 0.792 with two Dense layers and a dropout rate of 0.5, and BinaryNet achieved the best AUC of 0.79 with two Dense layers and a dropout rate of 0.6. Thus, the optimal dropout rate and the number of Dense layers for the two models were determined.

Figure 3(B) illustrates the architectures of three models with multi-head attention mechanism The number of attention heads and the dropout rate are the tuned hyperparameters. A series of attention heads and Dropout rate parameter settings were explored for the FusionCodonNet, HashNet, and PEBNet models for the optimal parameters, and the results of AUC are presented in Figure 5. A comparison reveals that FusionCodonNet and HashNet achieve the best recognition performance when the attention heads and Dropout rate are set to 0.3 and 8. Meanwhile, PEBNet exhibits the best recognition performance when the attention heads and Dropout rate are set to 0.4 and 4. Therefore, the Dropout rate and the number of attention heads for FusionCodonNet, HashNet, and PEBNet were determined.

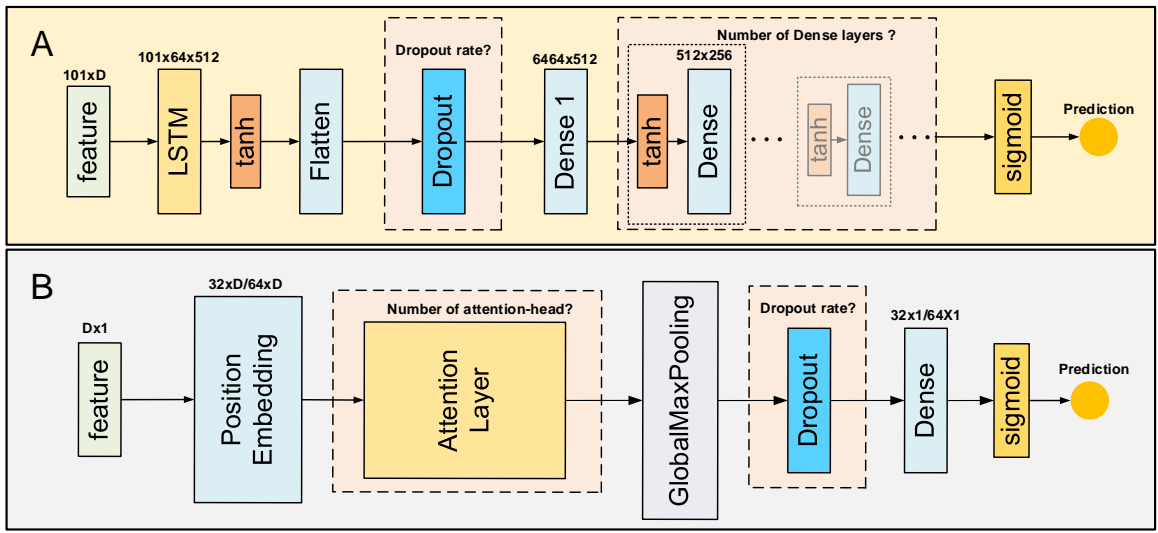

**Figure 3.** Structure and hyperparameter determination of the biological language models. (A) Structure of the LSTM-based models. (B) Structure of models with muti-head attention mechanism.

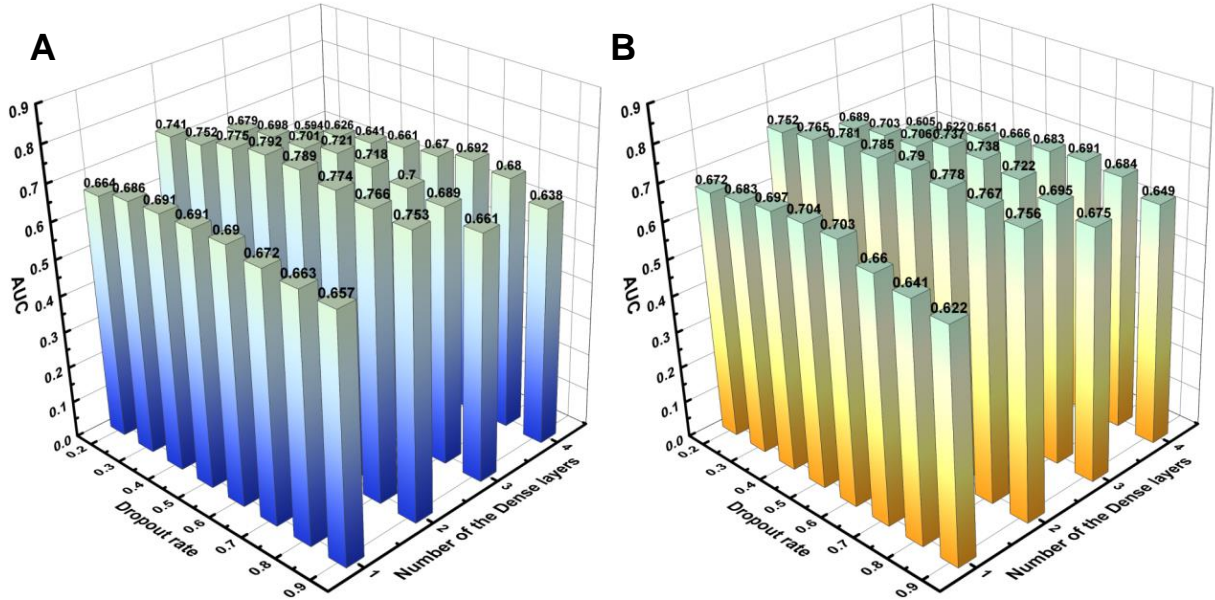

**Figure 4.** AUC of LSTM-based models with different hyperparameters. (A) OnehotNet with different Dropout rates and numbers of Dense layers. (B) BinaryNet with different Dropout rates and numbers of Dense layers.

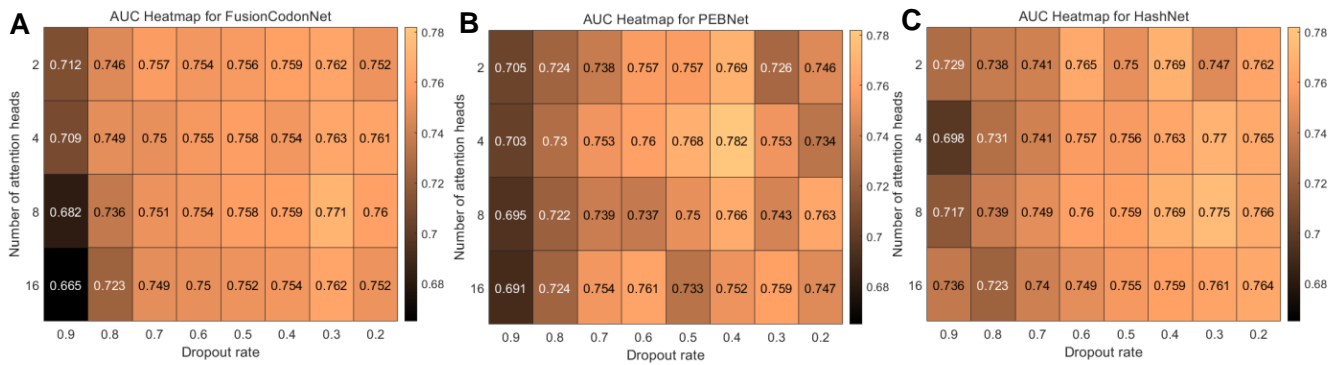

**Figure 5.** AUC of attention mechanism-based models with different hyperparameters. (A) FusionCodonNet with varying dropout rates and a range of attention head configurations. (B) HashNet with varying dropout rates and a range of attention head configurations. (C) PEBNet with varying dropout rates and a range of attention head configurations.

## *4.3 Performance comparison with different model architecture*

Utilizing the five sets of optimal hyperparameters obtained in section 4.2, f5C-finder was constructed by employing a linear ensemble method that integrates OnehotNet, BinaryNet, HashNet, PEBNet, and FusionCodonNet. To demonstrate the superior performance of the ensemble language model architecture in f5C identification, comparisons were made with ensemble models of three traditional architectures, they were SVM-based, ADA-based, and RF-based. For hyperparameter tuning, the AUC metric was also adopted, and Grid Search with 10-fold cross-validation strategy was utilized on the same training data. This process yielded AUC values for each sub-model with different architecture under various hyperparameter settings, as presented in Table 1. For the five SVM sub-classifiers, the radial basis function (rbf) kernel generally yielded the best performance for f5C modification recognition, and optimal results were achieved using a gamma value less than 0.055 and a C value greater than 3. For the hyperparameters of the five ADA sub-classifiers, the optimal numbers of weak learners and learning rates ranged from 140 to 190, and 0.001 to 0.01. For the five RF sub-classifiers, the optimal number and maximum depth of decision trees ranged from 350 to 450, and 37 to 46, resulting in the best f5C recognition performance. Utilizing the sub-models trained with these optimal parameters, the classifiers of the three architectures were linearly integrated to obtain Fusion-RF, Fusion-SVM, and Fusion-ADA, respectively.

Among the five single biological language models in Table 2, BinaryNet and OnehotNet exhibited superior performance in identifying f5C modifications. BinaryNet achieved Sn, Sp, MCC, F1, ACC, and AUC values of 0.742, 0.693, 0.436, 0.725, 0.718, and 0.790, respectively. OnehotNet yielded comparable results with corresponding values of 0.740, 0.695, 0.435, 0.723, 0.717, and 0.792. The three biological language models incorporating attention mechanisms demonstrated slightly lower ACC and AUC scores, ranging from 0.01 to 0.02 lower than BinaryNet and OnehotNet. The ensemble biological language model, f5C-finder, achieved the best results for SP, MCC, F1, ACC, and AUC, with values of 0.693, 0.445, 0.73, 0.722, and 0.807, respectively. Compared to Fusion-RF, Fusion-SVM, and Fusion-ADA, f5C-finder showed improvements ranging from 0.011 to 0.057 for SP, 0.034 to 0.082 for MCC, 0.011 to 0.049 for F1, 0.018 to 0.041 for ACC, and 0.038 to 0.06 for AUC. Therefore, f5C-finder which utilizes a biological language model architecture achieved the best results in 10-fold cross-validation for the f5C modification identification.

**Table 1** Hyperparameter optimization results of SVM, ADA, and RF.

| Model | Hyperparameters | Meaning | Search ranges | Optimal values of each feature | | | | |
|---|---|---|---|---|---|---|---|---|
| | | | | Onehot | Binary | Hash | PEB | FusionCodon |
| SVM | C | Regularization parameter | (0.1, 50) | 10.655 | 8.795 | 3.680 | 13.250 | 6.845 |
| | gamma | Kernel coefficient | (0.001, 1) | 0.002 | 0.010 | 0.055 | 0.005 | 0.045 |
| | kernel | Type of kernel function mapping data into higher dimensional space | (linear, rbf, poly, sigmoid) | rbf | rbf | rbf | rbf | rbf |
| ADA | n_estimators | Number of weak learners | (10, 200) | 180 | 160 | 180 | 140 | 190 |
| | learning_rate | Learning rate | (0.0001, 1) | 0.001 | 0.001 | 0.01 | 0.01 | 0.001 |
| RF | n_estimators | Number of decision trees | (50, 500) | 400 | 450 | 350 | 400 | 450 |
| | max_deep | Maximum depth of each decision tree | (2, 50) | 42 | 39 | 37 | 43 | 46 |

**Note:** the details of the hyperparameters can be referenced from https://scikit-learn.org/stable/

**Table 2** Model performances on 10-fold cross-validation.

| Type | Model | SN | SP | MCC | F1 | ACC | AUC |
|---|---|---|---|---|---|---|---|
| Single-model | OnehotNet | 0.740 | **0.695** | 0.435 | 0.723 | 0.717 | **0.792** |
| | BinaryNet | 0.742 | 0.693 | **0.436** | **0.725** | **0.718** | 0.790 |
| | FusionCodonNet | 0.734 | 0.672 | 0.407 | 0.712 | 0.703 | 0.771 |
| | HashNet | **0.749** | 0.668 | 0.418 | 0.719 | 0.708 | 0.775 |
| | PEBNet | 0.732 | 0.692 | 0.424 | 0.718 | 0.712 | 0.782 |
| Ensemble model | Fusion-RF | **0.758** | 0.651 | 0.411 | 0.719 | 0.704 | 0.769 |
| | Fusion-SVM | 0.681 | 0.682 | 0.363 | 0.681 | 0.681 | 0.747 |
| | Fusion-ADA | 0.749 | 0.636 | 0.388 | 0.709 | 0.693 | 0.754 |
| | f5C-finder | 0.751 | **0.693** | **0.445** | **0.730** | **0.722** | **0.807** |

## 4.4. Model performance on the independent test

To further validate the advantages of the ensemble biological language models for the f5C modification recognition, independent test was conducted. Additionally, the modeling capability of the biological language models was analyzed based on the principles of the multi-head attention mechanism, and it provides insights into its ability to capture relevant patterns and dependencies within RNA sequences. This analysis demonstrates the superiority of language model-based ensemble classifiers for f5C modification recognition and elucidates the underlying mechanisms contributing to their enhanced performance. Table 3 presents the performance of different models across the six metrics on independent test. Compared with the five single models, BinaryNet exhibited the best performance on the five metrics SP, MCC, F1, ACC, and AUC, with values of 0.767, 0.496, 0.743, 0.748, and 0.815, respectively. It is worth noting that BinaryNet outperformed Fusion-RF, Fusion-SVM, and Fusion-ADA across the 5 metrics, by 0.127 to 0.147, 0.123 to 0.139, 0.044 to 0.055, 0.062 to 0.071, and 0.054 to 0.077. This suggests that language model architecture has a clear advantage in the recognition of f5C. Among the four fusion models, f5C-finder achieved the best results for all six metrics, with values of 0.737, 0.737, 0.474, 0.737, 0.737, and 0.827. outperforms Fusion-RF, Fusion-SVM, and Fusion-ADA across the 6 metrics, by 0.003 to 0.044, 0.058 to 0.117, 0.101 to 0.117, 0.038 to 0.049, 0.05 to 0.061, and 0.066 to 0.077. Therefore, f5C-finder is the best model for the identification of the f5C modification with the highest accuracy and robustness.

**Table 3** Model comparison on the independent test dataset.

| Type | Model | SN | SP | MCC | F1 | ACC | AUC |
| --- | --- | --- | --- | --- | --- | --- | --- |
| **Single-model** | **OnehotNet** | **0.776** | 0.684 | 0.462 | 0.742 | 0.730 | 0.804 |
| | **BinaryNet** | 0.729 | **0.767** | **0.496** | **0.743** | **0.748** | **0.815** |
| | **FusionCodonNet** | 0.770 | 0.684 | 0.456 | 0.738 | 0.727 | 0.783 |
| | **HashNet** | 0.737 | 0.706 | 0.443 | 0.726 | 0.722 | 0.790 |
| | **PEBNet** | 0.765 | 0.701 | 0.466 | 0.741 | 0.733 | 0.793 |
| **Ensemble model** | **Fusion-RF** | 0.731 | 0.640 | 0.373 | 0.699 | 0.686 | 0.761 |
| | **Fusion-SVM** | 0.693 | 0.679 | 0.371 | 0.688 | 0.686 | 0.753 |

| | | | | | | |
|---|---|---|---|---|---|---|
| Fusion-ADA | 0.734 | 0.620 | 0.357 | 0.695 | 0.677 | 0.750 |
| f5C-finder | **0.737** | **0.737** | **0.474** | **0.737** | **0.737** | **0.827** |

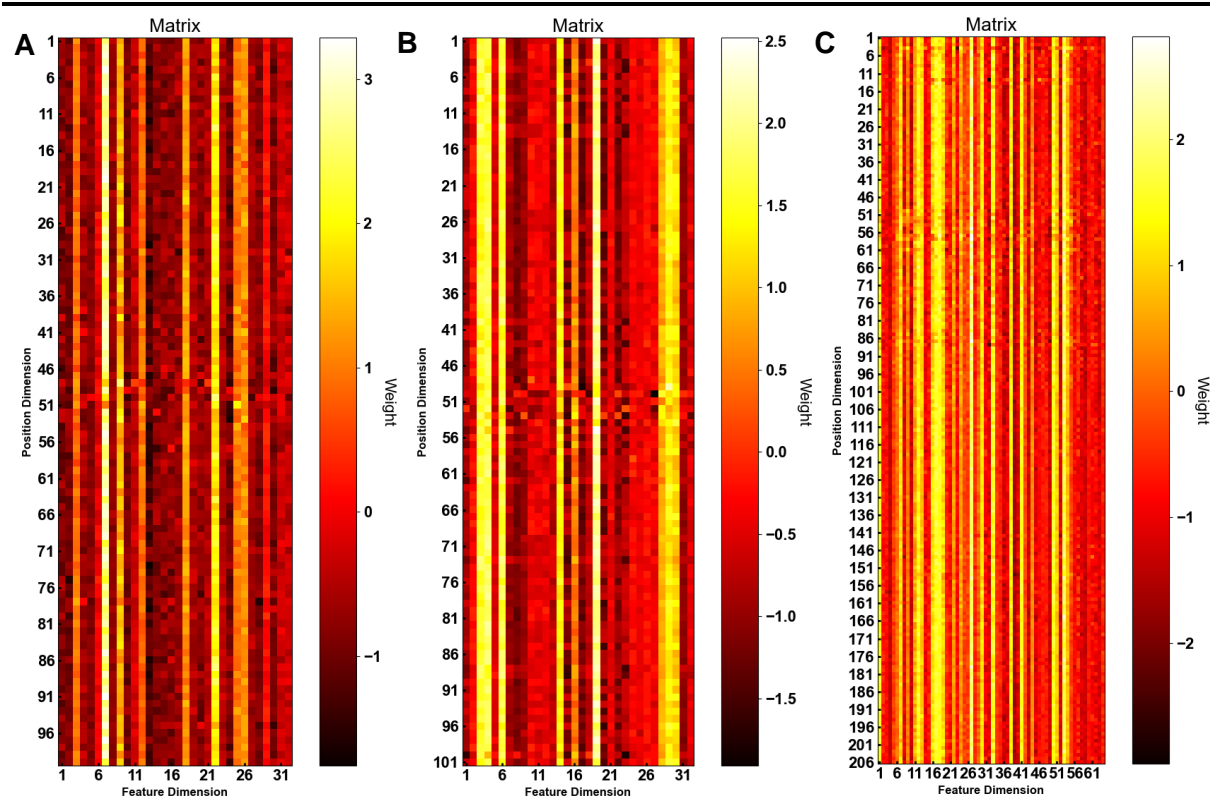

**Figure 6.** Visualization of output matrices from biological language model layers. Here, matrices are the inference results of the same sample with f5C modification. (A) Attention core output in HashNet. (B) Attention core output in PEBNet. (C) Attention core output in FusionCodonNet. Note: the horizontal axis represents the feature dimensions of the output, the vertical axis represents the dimensions of positional information, and the color in the heatmap represents the modeling results of the model for the input sequence.

To elucidate the knowledge acquired by biological language models during training, a visual representation of the output matrices from the trained language model layers (Attention core output) is illustrated in Figure 6. This visualization demonstrates the effectiveness of incorporating position-based feature extraction and attention mechanisms in modeling RNA sequences containing the f5C modification. Firstly, for the various position-based feature extraction methods in this paper, slicing the output matrix parallel from the feature axis, the obtained vectors have strong similarities. The colors of these vectors are roughly similar, indicating that each vector still maintains information about the corresponding A, G, C, U. And the slight color differences of these vectors contain information about the four nucleic acid types and their positions. This indicates these language models can indeed leverage positional information for sequence modeling. Secondly, the use of attention mechanisms plays a certain role. In Figure 6(A) and Figure 6(B), the matrix values around the cytosine at position 51 show irregular and dramatic changes in color depth, corresponding to the distribution of A, G, C, and T around the cytosine in Figure 2. This suggests that the language models can capture the information-rich region of nucleotides, pay attention to the information around the cytosine which is more significant, and map the input vector into one with global information (relationship between nucleotide). Therefore, the utilization of attention mechanisms is effective. Finally, this explains the superiority of the language model architecture over tree classifier architecture (RF), linear classifier architecture (SVM), and adaptive boosting architecture (ADA) for modification identification. The ability to model sequence positional information and simultaneously notice effective positions is something that these three traditional architecture models lack.

### *4.5. Model comparison with SOTA models*

F5C-finder is the first machine learning model designed for f5C modification identification. To validate its superior performance, three established ensemble models known for their high accuracy in identifying the m7G, 6mA, and m5C modifications were trained on the f5C dataset, they are BERT-m7G, i6mA-Vote, and m5C-HPromoter. These models represent diverse architectures, providing a robust basis for comparison. BERT-m7G utilizes the BERT language model with feature selection, i6mA-Vote integrates Onehot features and five sub-models including neural network-based multi-layer perceptron and statistical learning classifiers, while m5C-HPromoter adopts Onehot features and a stacking-based ensemble deep learning classifier. This diversity in model architectures provides a meaningful basis for comparison, as each model offers unique strengths for f5C identification. To ensure a fair comparison, the same data split was used for model training in the 10-fold cross-validation. For the training of the three models, the feature optimization or model parameter tuning processes mentioned in the original papers were conducted on the f5C training data to achieve optimal results. As shown in Figure 7, f5C-finder achieved the highest prediction performance across five metrics: SN, MCC, F1, ACC, and AUC on 10-fold cross-validation. Compared to i6mA-Vote, f5C-finder exhibited improvements of 0.112, 0.044, 0.05, 0.023, and 0.033 in these respective metrics, with only a lower value of 0.066 in SP. However, the superiority in SN achieved by f5C-finder outweighed the decrease in SP, and f5C-finder surpassed i6mA-Vote in both ACC and MCC. These results demonstrated the superior performance of f5C-finder in identifying the f5C modifications. Additionally, f5C-finder outperformed both BERT-m7G and m5C-HPromoter across all six metrics, with improvements of 0.035, 0.007, 0.043, 0.025, 0.021, and 0.038 for BERT-m7G and 0.04, 0.009, 0.051, 0.029, 0.025, and 0.069 for m5C-HPromoter in SN, SP, MCC, F1, ACC, and AUC, respectively. Therefore, f5C-finder is the best model for f5C identification.

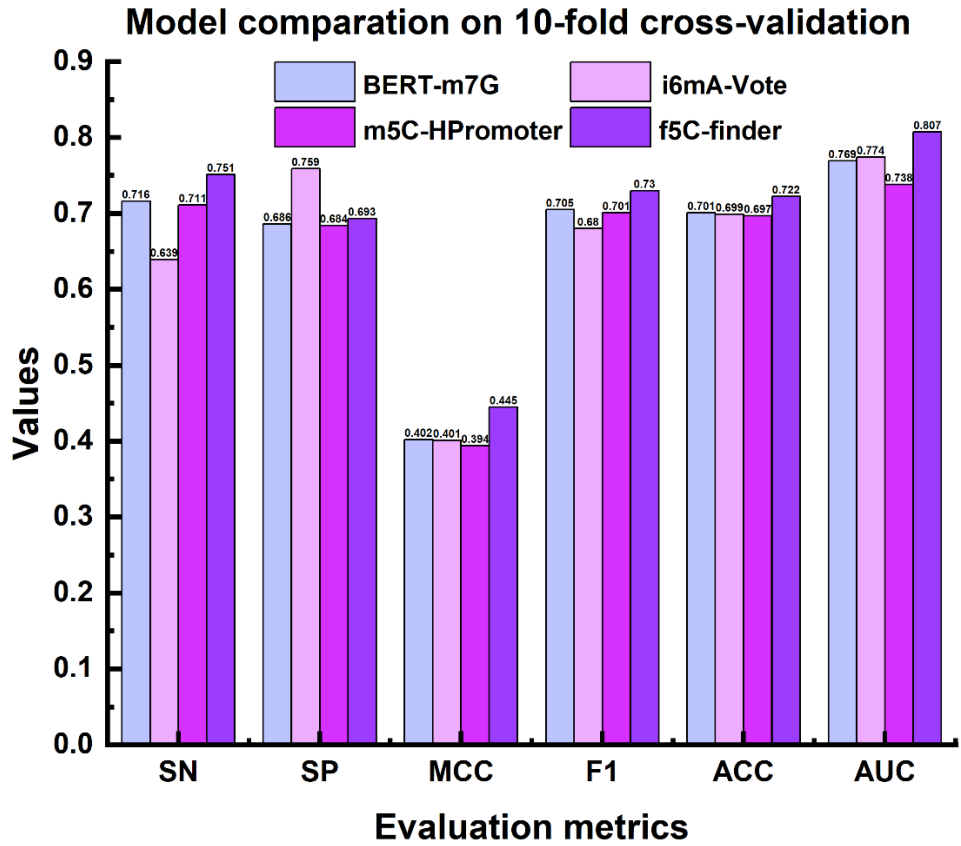

**Figure 7.** Evaluation metrics of different models for f5C identification on 10-fold cross-validation.

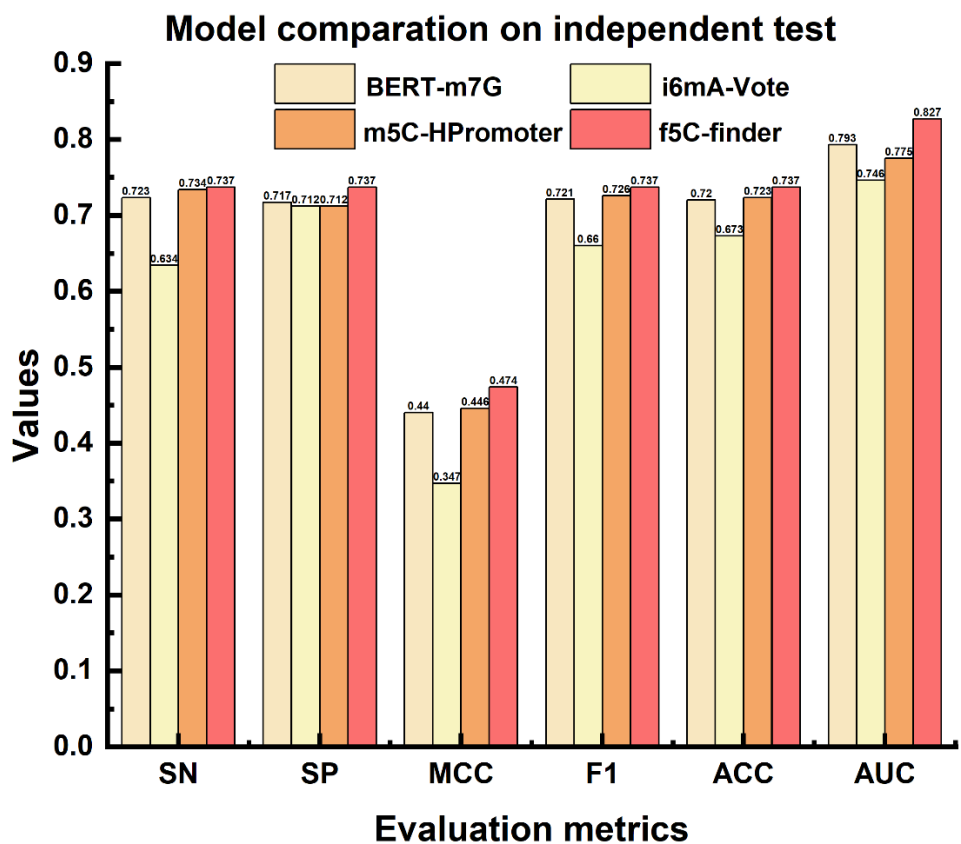

**Figure 8.** Evaluation metrics of different models for f5C identification on independent test.

Additionally, for the independent test, the entire training dataset was utilized for training, and the six metrics were employed for performance evaluation on the test set. The results were illustrated in Figure 8. F5C-finder achieved the highest values across all six metrics in the independent test, it outperformed BERT-m7G, i6mA-Vote, and m5C-HPromoter by 0.016, 0.077, and 0.011 in F1, respectively, and by 0.017, 0.064, and 0.014 in ACC, respectively. And f5C-finder exhibited even more pronounced advantages on the AUC and MCC metrics. It outperformed the three ensemble models, BERT-m7G, i6mA-Vote, and m5C-HPromoter, by 0.034, 0.081, and 0.052 in AUC, and by 0.034, 0.127, and 0.028 in MCC, respectively. Therefore, f5C-finder, an ensemble classifier based on a language model architecture, is the SOTA model for f5C identification.

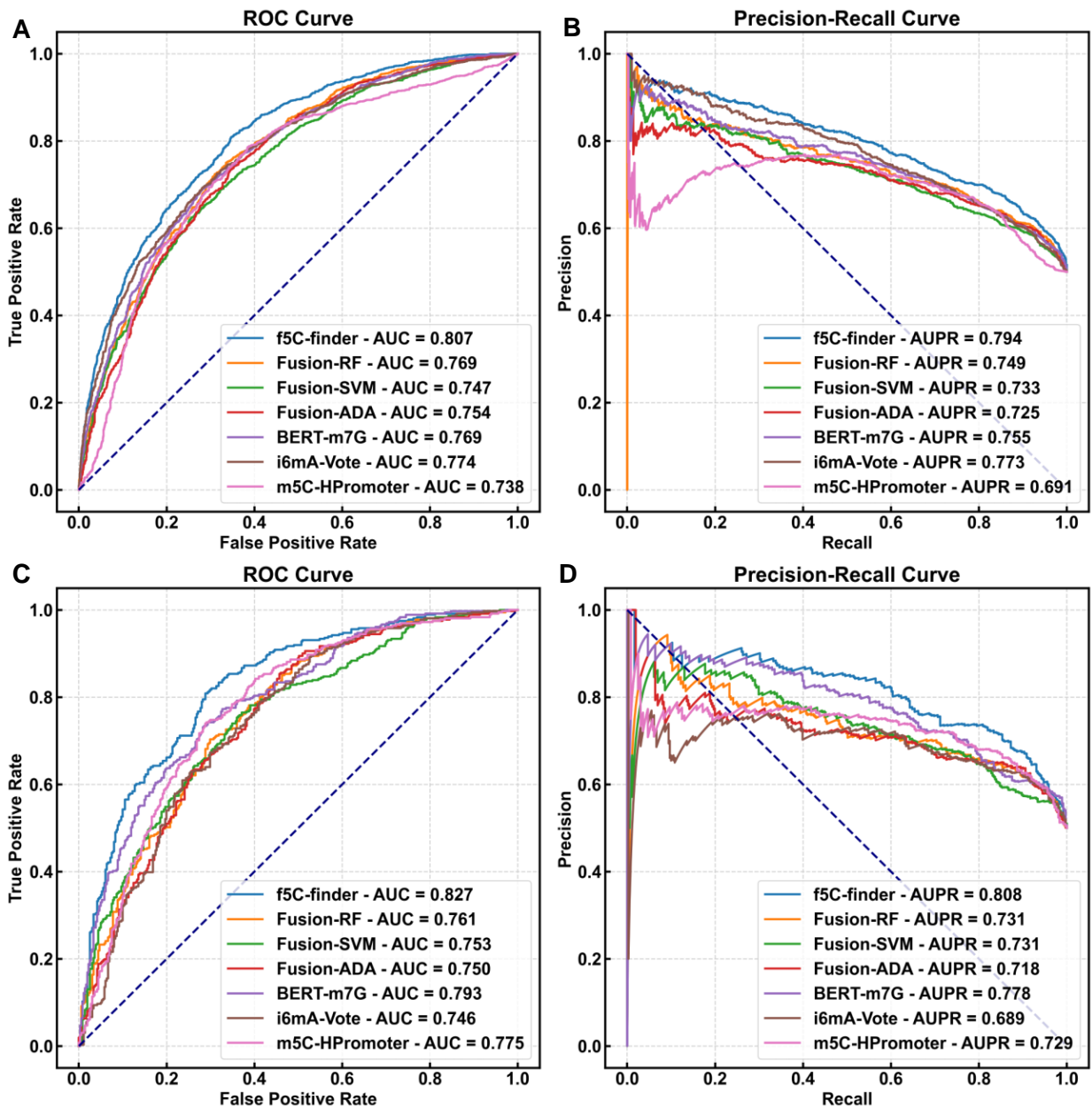

**Figure 9.** ROC Curves and P-R Curves of different models. (A) ROC Curves of the models on 10-fold cross-validation. (B) P-R Curves of the models on 10-fold cross-validation. (C) ROC Curves of the models on independent test. (D) P-R Curves of the models on independent test.

To validate the robustness of f5C-finder, the ROC Curves and P-R Curves of the ensemble models were further illustrated. The results from 10-fold cross-validation are shown in Figure 9(A) and Figure 9(B). The area under the ROC curve (AUC) for f5C-finder was consistently larger than that of Fusion-RF, Fusion-ADA, Fusion-SVM, BERT-m7G, i6mA-Vote, and m5C-HPromoter, with improvement ranging from 0.033 to 0.069. Similarly, the area under the P-R curve (AUPR) for f5C-finder exceeded the other models by about 0.021 to 0.103. Figure 9(C) and 9(D) illustrate the results from the independent test, f5C-finder demonstrated superior performance with AUC values exceeding the other models by 0.034 to 0.081 and area under the P-R curve surpassing them by 0.03 to 0.119. These findings demonstrated the superior accuracy and robustness of f5C-finder for the f5C modification identification. Analyzing the shape of the curves provides insights into the dynamic relationship between TP and FP, as well as the dynamic relationship between Precision and Recall, allowing for a comparison of model robustness. Whether in 10-fold cross-validation or independent testing, the curves corresponding to f5C-finder have relatively fewer spikes and smoother changes, indicating better robustness of f5C-finder model. The superior performance of f5C-finder in both 10-fold cross-validation and independent tests demonstrated the advantages of ensemble learning classifiers with complete language model architectures for the f5C modification identification. While BERT-m7G utilized the BERT language model architecture, the single feature scheme (Onehot) limited the model ability to fully capture the complexity information of RNA sequences, and this resulted in the inferior performance compared to f5C-finder. While i6mA-Vote integrated five different machine learning classifiers, the lack of neural network-based language models hindered its ability to fully capture the deeper information of the f5C modifications. This resulted in its inferior performance for f5C identification compared to f5C-finder and highlighted the advantages of language model architecture in RNA information modeling. Similarly, m5C-HPromoter, despite employing a stacking-based ensemble deep learning classifier and achieving better performance than i6mA-Vote in several metrics, it still underperformed compared to f5C-finder. This further underscored the strength of ensemble biological language model for f5C identification.

## 5. Webserver implementation

A powerful tool can provide great convenience for researchers who lack a foundation in mathematics or computer science. Therefore, this study developed an online prediction platform for f5C identification based on f5C-finder. This platform enables users to utilize the model directly without requiring extensive computational expertise. It offers various functionalities, including introduction, prediction, download, and contact options. As depicted in Figure 10, users can navigate to the 'Server' page, input a 101-base pair (bp) sequence in FASTA format into the designated text box, and click 'Submit' to determine the presence of f5C modification. For clarity on the data input format, clicking the 'Example(s)' button displays sample RNA sequences in the text box. The 'Data' page provides access to the divided dataset for download. Any inquiries regarding the website can be directed to the author through the 'Contact' page.

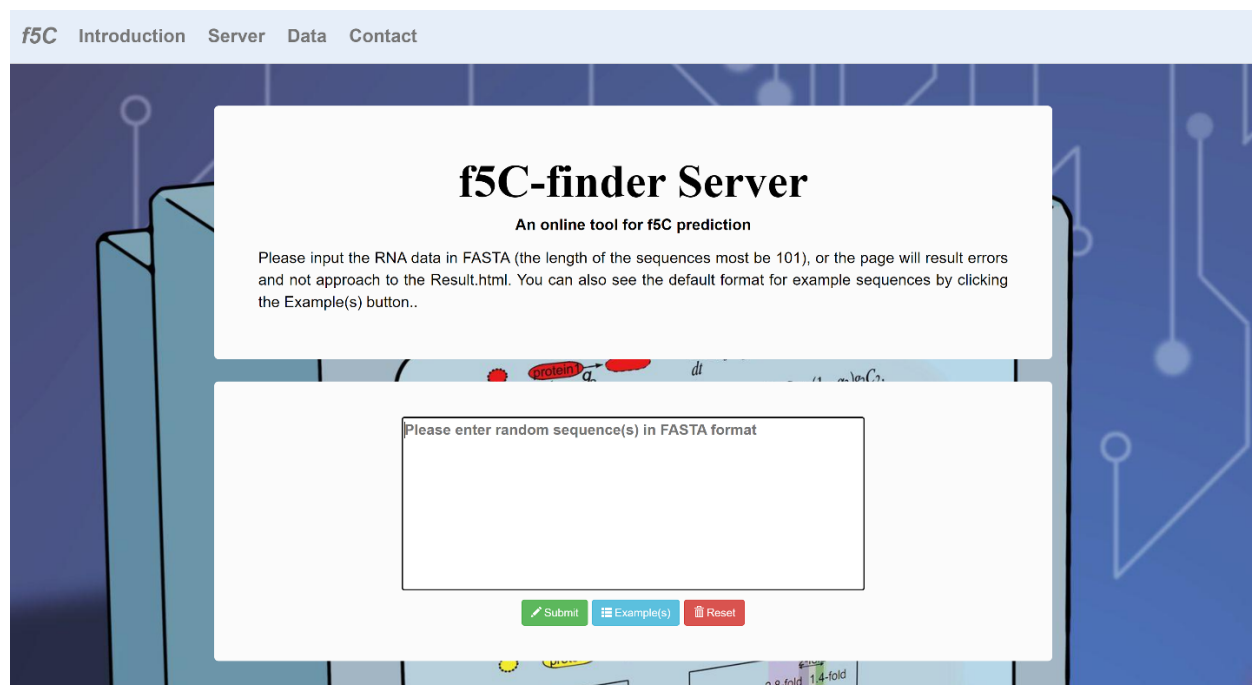

**Figure 10.** Webserver of f5C-finder.

## 6. Conclusion

This study introduced f5C-finder, the first ensemble biological language model for identifying RNA f5C modifications. Through rigorous validation and testing using six performance metrics and ROC/P-R curves, f5C-finder consistently outperformed existing ensemble models, including BERT-m7G, i6mA-Vote, and m5C-HPromoter, it achieved AUC values that were 0.033 to 0.069 and 0.034 to 0.081 higher than these three ensemble models in 10-fold cross-validation and independent test, respectively. These results demonstrated that f5C-finder is the SOTA model for the f5C modification identification.

During the process, deep biological language learning model structure has demonstrated with better identification performances than the classic statistical machine learning structures. This result highlighted the strength of the deep language model in capturing both the sequential order and functional meaning within genomic sequences. Furthermore, the integrated interpretable analysis mechanism provides valuable insights into the model's learning process. This allows us to connect the identification of key sequential features to a deeper understanding of their biological functions.

However, there is significant potential for further improvement. One limitation of the current model is its focus on analyzing only local sequences surrounding the f5C sites. This approach might restrict the capture of all relevant discriminative information. At the same time, due to the limited number of samples in the dataset, the deep biological language learning model structure cannot be fully utilized. Therefore, in the future, more data will be collected, and the model's identification performance for the f5C modification will be further improved.

**Data availability**

The dataset used in this study and the source code of f5C-finder are all available at https://github.com/NWAFU-LiuLab/ f5C-finder.

**CRediT authorship contribution statement**

**Guohao Wang**: Formal analysis, Investigation, Methodology, Visualization, Writing - original draft, review & editing. **Ting Liu**: Methodology, Software. **Hongqiang Lyu:** Conceptualization, Writing - review & editing. **Ze Liu:** Conceptualization, Resources, Funding acquisition, Supervision, Writing - review & editing.

**Declaration of competing interest**

The authors declare that they have no known competing financial interests or personal relationships that could have appeared to influence the work reported in this paper.

**Funding**

This work was supported by the National Natural Science Foundation of China (61902323), and the Start-up foundation of Northwest A&F University (Z109021809).

**References**

[1]     Moriya, J. et al. (1994). A novel modified nucleoside found at the first position of the anticodon of methionine tRNA from bovine liver mitochondria. *Biochemistry*. DOI: 10.1021/bi00174a033.

[2]     Watanabe, Y. et al. (1994). Primary and higher order structures of nematode (Ascaris suum) mitochondrial tRNAs lacking either the T or D stem. *The Journal of Biological Chemistry*.

[3]     Tomita, K. et al. (1997). 5-formylcytidine (f5C) found at the wobble position of the anticodon of squid mitochondrial tRNA(Met)CAU. *Nucleic Acids Symposium Series*.

[4]     Takemoto, C. et al. (1999). Nucleotide sequences of animal mitochondrial tRNAs(Met) possibly recognizing both AUG and AUA codons. *Nucleic Acids Symposium Series*. DOI: 10.1093/nass/42.1.77.

[5]     Tomita, K. et al. (1999). Codon reading patterns in Drosophila melanogaster mitochondria based on their tRNA sequences: a unique wobble rule in animal mitochondria. *Nucleic Acids Research*. DOI: 10.1093/nar/27.21.4291.

[6]     Van Haute, L. et al. (2016). Deficient methylation and formylation of mt-tRNA(Met) wobble cytosine in a patient carrying mutations in NSUN3. *Nature Communications*. DOI: 10.1038/ncomms12039.

[7]     Nakano, S. et al. (2016). NSUN3 methylase initiates 5-formylcytidine biogenesis in human mitochondrial tRNA(Met). *Nature Chemical Biology*. DOI: 10.1038/nchembio.2099.

[8]     Takemoto, C. et al. (2009). Unconventional decoding of the AUA codon as methionine by mitochondrial tRNAMet with the anticodon f5CAU as revealed with a mitochondrial in vitro translation system. *Nucleic Acids Research*. DOI: 10.1093/nar/gkp001.

[9]     Murakami, Y. et al. (2023). NSUN3-mediated mitochondrial tRNA 5-formylcytidine modification is essential for embryonic development and respiratory complexes in mice. *Communications Biology*. DOI: 10.1038/s42003-023-04680-x.

[10]    Lusic, H. et al. (2008). Synthesis and investigation of the 5-formylcytidine modified, anticodon stem and loop of the human mitochondrial tRNAMet. *Nucleic Acids Research*. DOI: 10.1093/nar/gkn703.

[11]    Xia, B. et al. (2015). Bisulfite-free, base-resolution analysis of 5-formylcytosine at the genome scale. *Nature Methods*. DOI: 10.1038/nmeth.3569.

[12]    C, Z. et al. (2017). Single-Cell 5-Formylcytosine Landscapes of Mammalian Early Embryos and ESCs at Single-Base Resolution. *Cell stem cell*. DOI: 10.1016/j.stem.2017.02.013.

[13]    Lu, X. et al. (2015). TET family proteins: oxidation activity, interacting molecules, and functions in diseases. *Chemical Reviews*. DOI: 10.1021/cr500470n.

[14]    Liu, Y. et al. (2019). Bisulfite-free direct detection of 5-methylcytosine and 5-hydroxymethylcytosine at base resolution. *Nature Biotechnology*. DOI: 10.1038/s41587-019-0041-2.

[15]    Habibi, E. et al. (2013). Whole-genome bisulfite sequencing of two distinct interconvertible DNA methylomes of mouse embryonic stem cells. *Cell Stem Cell*. DOI: 10.1016/j.stem.2013.06.002.

[16]    Wang, Y. et al. (2022). Single-Base Resolution Mapping Reveals Distinct 5-Formylcytidine in *Saccharomyces cerevisiae* mRNAs. *ACS Chemical Biology*. DOI: 10.1021/acschembio.1c00633.

[17]    Lv, H. et al. (2019). iDNA6mA-Rice: A Computational Tool for Detecting N6-Methyladenine Sites in Rice. *Frontiers in Genetics*. DOI: 10.3389/fgene.2019.00793.

[18]    Feng, P. et al. (2019). iDNA6mA-PseKNC: Identifying DNA N6-methyladenosine sites by incorporating nucleotide physicochemical properties into PseKNC. *Genomics*. DOI: 10.1016/j.ygeno.2018.01.005.

[19]    Hasan, M.M. et al. (2020). i6mA-Fuse: improved and robust prediction of DNA 6 mA sites in the Rosaceae genome by fusing multiple feature representation. *Plant Molecular Biology*. DOI: 10.1007/s11103-020-00988-y.

[20]     Hasan, M.M. et al. (2021). Meta-i6mA: an interspecies predictor for identifying DNA *N* 6-methyladenine sites of plant genomes by exploiting informative features in an integrative machine-learning framework. *Briefings in Bioinformatics*. DOI: 10.1093/bib/bbaa202.

[21]     Teng, Z. et al. (2022). i6mA-Vote: Cross-Species Identification of DNA N6-Methyladenine Sites in Plant Genomes Based on Ensemble Learning With Voting. *Frontiers in Plant Science*. DOI: 10.3389/fpls.2022.845835.

[22]     Chen, W. et al. (2019). iRNA-m7G: Identifying N7-methylguanosine Sites by Fusing Multiple Features. *Molecular Therapy. Nucleic Acids*. DOI: 10.1016/j.omtn.2019.08.022.

[23]     Zhang, L. et al. (2021). BERT-m7G: A Transformer Architecture Based on BERT and Stacking Ensemble to Identify RNA N7-Methylguanosine Sites from Sequence Information. *Computational and Mathematical Methods in Medicine*. DOI: 10.1155/2021/7764764.

[24]     Chai, D. et al. (2021). Staem5: A novel computational approach for accurate prediction of m5C site. *Molecular Therapy. Nucleic Acids*. DOI: 10.1016/j.omtn.2021.10.012.

[25]     Xiao, X. et al. (2022). m5C-HPromoter: An Ensemble Deep Learning Predictor for Identifying 5-methylcytosine Sites in Human Promoters. *Current Bioinformatics*. DOI: 10.2174/1574893617666220330150259.

[26]     Radford, A. et al. Improving Language Understanding by Generative Pre-Training.

[27]     Brown, T. et al. (2020). Language Models are Few-Shot Learners. *Advances in Neural Information Processing Systems*.

[28]     Bai, Y. et al. (2022). Training a Helpful and Harmless Assistant with Reinforcement Learning from Human Feedback. DOI: 10.48550/arXiv.2204.05862.

[29]     Devlin, J. et al. (2019). BERT: Pre-training of Deep Bidirectional Transformers for Language Understanding. *Proceedings of the 2019 Conference of the North American Chapter of the Association for Computational Linguistics: Human Language Technologies, Volume 1 (Long and Short Papers)*. DOI: 10.18653/v1/N19-1423.

[30]     Hochreiter, S. and Schmidhuber, J. (1997). Long Short-Term Memory. *Neural Computation*. DOI: 10.1162/neco.1997.9.8.1735.

[31]     Sherstinsky, A. (2020). Fundamentals of Recurrent Neural Network (RNN) and Long Short-Term Memory (LSTM) Network. *Physica D: Nonlinear Phenomena*. DOI: 10.1016/j.physd.2019.132306.

[32]     Vaswani, A. et al. (2017). Attention is All you Need. *Advances in Neural Information Processing Systems*.

[33]     Zhao, H. et al. (2024). Explainability for Large Language Models: A Survey. *ACM Transactions on Intelligent Systems and Technology*. DOI: 10.1145/3639372.

[34] Li, C.-C. and Liu, B. (2020). MotifCNN-fold: protein fold recognition based on fold-specific features extracted by motif-based convolutional neural networks. *Briefings in Bioinformatics*. DOI: 10.1093/bib/bbz133.

[35] Wei, L. et al. (2018). Prediction of human protein subcellular localization using deep learning. *Journal of Parallel and Distributed Computing*.

[36] Wang, Y. et al. (2022). Effector-GAN: prediction of fungal effector proteins based on pretrained deep representation learning methods and generative adversarial networks. *Bioinformatics (Oxford, England)*. DOI: 10.1093/bioinformatics/btac374.

[37] Elnaggar, A. et al. (2021). ProtTrans: Toward Understanding the Language of Life Through Self-Supervised Learning. *MACHINE INTELLIGENCE*.

[38] Meng, J. et al. (2021). PlncRNA-HDeep: plant long noncoding RNA prediction using hybrid deep learning based on two encoding styles. *BMC bioinformatics*. DOI: 10.1186/s12859-020-03870-2.

[39] Wang, H. et al. (2022). EMDLP: Ensemble multiscale deep learning model for RNA methylation site prediction. *BMC bioinformatics*. DOI: 10.1186/s12859-022-04756-1.

[40] Breiman, L. (2001). Random Forests. *Machine Learning*. DOI: 10.1023/A:1010933404324.

[41] Ao, C. et al. (2023). m5U-SVM: identification of RNA 5-methyluridine modification sites based on multi-view features of physicochemical features and distributed representation. *BMC Biology*. DOI: 10.1186/s12915-023-01596-0.

[42] Javeed, A. et al. (2022). An Intelligent Learning System for Unbiased Prediction of Dementia Based on Autoencoder and Adaboost Ensemble Learning. *Life*. DOI: 10.3390/life12071097.

[43] Li, W. and Godzik, A. (2006). Cd-hit: a fast program for clustering and comparing large sets of protein or nucleotide sequences. *Bioinformatics (Oxford, England)*. DOI: 10.1093/bioinformatics/btl158.

[44] Dao, F.-Y. et al. (2021). A computational platform to identify origins of replication sites in eukaryotes. *Briefings in Bioinformatics*. DOI: 10.1093/bib/bbaa017.

[45] Misra, S. et al. (2011). Anatomy of a hash-based long read sequence mapping algorithm for next generation DNA sequencing. *Bioinformatics*. DOI: 10.1093/bioinformatics/btq648.

[46] Abbas, Z. et al. (2022). ENet-6mA: Identification of 6mA Modification Sites in Plant Genomes Using ElasticNet and Neural Networks. *International Journal of Molecular Sciences*. DOI: 10.3390/ijms23158314.

[47] Chen, P.-C. et al. (2021). A Simple and Effective Positional Encoding for Transformers. *Proceedings of the 2021 Conference on Empirical Methods in Natural Language Processing*. DOI: 10.18653/v1/2021.emnlp-main.236.

[48]     Nabeel Asim, M. et al. (2023). DNA-MP: a generalized DNA modifications predictor for multiple species based on powerful sequence encoding method. *Briefings in Bioinformatics*. DOI: 10.1093/bib/bbac546.

[49]     Pang, C. et al. (2023). Advanced deep learning methods for molecular property prediction. *Quantitative Biology*. DOI: 10.1002/qub2.23.

[50]     Fang, Y. et al. (2023). AFP-MFL: accurate identification of antifungal peptides using multi-view feature learning. *Briefings in Bioinformatics*. DOI: 10.1093/bib/bbac606.

[51]     Jiang, Y. et al. (2023). Explainable Deep Hypergraph Learning Modeling the Peptide Secondary Structure Prediction. *Advanced Science*. DOI: 10.1002/advs.202206151.

[52]     Luo, Z. et al. (2022). Predicting N6-Methyladenosine Sites in Multiple Tissues of Mammals through Ensemble Deep Learning. *International Journal of Molecular Sciences*. DOI: 10.3390/ijms232415490.